\documentclass[reprint,prb,aps,twocolumn,superscriptaddress,10pt]{revtex4-2}
\usepackage{amsmath}
\usepackage[english]{babel}
\usepackage{graphicx}
\usepackage{color}
\usepackage{sidecap}
\usepackage[dvipsnames]{xcolor}

\usepackage[colorlinks,
            citecolor = blue, 
						linkcolor = blue,
						urlcolor  = blue,
						anchorcolor = blue,
						breaklinks = true
						]{hyperref}

\usepackage{cleveref}

\usepackage{soul}
\usepackage{ amssymb }

\newif\ifdraft
\drafttrue

\def \ETH{Institute for Quantum Electronics, ETH Z\"urich, CH-8093 Z\"urich, Switzerland}
\def \UW{Department of Physics, University of Washington, Seattle, WA, USA}
\def \NIMSKW{Research Center for Electronic and Optical Materials, National Institute for Materials Science, 1-1 Namiki, Tsukuba 305-0044, Japan}
\def \NIMSTT{Research Center for Materials Nanoarchitectonics, National Institute for Materials Science,  1-1 Namiki, Tsukuba 305-0044, Japan}

\def \Basel{Department of Physics, University of Basel, Klingelbergstrasse 82, Basel CH-4056, Switzerland}

\begin{document}


\title{Optical control over topological Chern number in moir\'e materials}

\author{O. Huber}
\affiliation{\ETH}

\author{K. Kuhlbrodt}
\affiliation{\ETH}

\author{E. Anderson}
\affiliation{\UW}

\author{W. Li}
\affiliation{\UW}

\author{K.~Watanabe}
\affiliation{\NIMSKW}

\author{T.~Taniguchi}
\affiliation{\NIMSTT}

\author{M. Kroner}
\affiliation{\ETH}

\author{X. Xu}
\affiliation{\UW}

\author{A. Imamo\u{g}lu}
\email{imamoglu@ethz.ch}
\affiliation{\ETH}

\author{T. Smole\'nski}
\email{tomasz.smolenski@unibas.ch}
\affiliation{\Basel}

\maketitle

{\bf 
Controlling quantum matter with light offers a promising route to dynamically tune its many-body properties, ranging from band topology~\cite{Wang_Science_2013,Mitra_Nature_2024} to superconductivity~\cite{Fausti_2011, Mitrano_2016,Rowe_2023}. However, achieving such optical control for strongly correlated electron systems in the steady-state has remained elusive. Here, we demonstrate all-optical switching of the spin-valley degree of freedom of itinerant ferromagnets in twisted MoTe$_2$ homobilayers. This system uniquely features flat valley-contrasting Chern bands and exhibits a range of strongly correlated phases at various moiré lattice fillings, including Chern insulators and ferromagnetic metals~\cite{Cai_Xu_2023,Zeng_Nature_2023,Xu_PRX_2023,Park_Xu_2023}. We show that the spin-valley orientation of all of these phases can be dynamically reversed by resonantly exciting the attractive polaron transition with circularly-polarized light. These findings not only constitute the first direct evidence for non-thermal switching of a ferromagnetic spin state at zero magnetic field, but also demonstrate the possibility of dynamical control over topological-order parameter, paving the way for all-optical generation of chiral edge modes and topological quantum circuits.
}



The ground states of physical systems in nature usually consist of multiple energy levels that are nearly-degenerate with respect to internal degrees of freedom, such as spin or orbital angular momentum. A unique approach allowing for dynamical initialization of such systems is optical orientation -- a process where light with a given polarization is used to drive the system to a selected metastable ground state that persists long after the excitation is turned off. Over the past few decades, this method has proven to be an invaluable tool enabling ultrafast control over a wide selection of systems, ranging from atomic ensembles~\cite{Kastler_1950,Kastler_Brossel_Winter_1952}, through spinful dopants in semiconductors~\cite{Zacharchenya_Meier_book}, to solid-state nuclear spin systems~\cite{Awschalom_2013,Atature_2019}. Recently, a theory proposal indicated that optical pumping could be used to control topological states~\cite{Victor_PRB}.

Despite these advances, optical control over correlated many-body systems represents an outstanding challenge. Already for bulk ferromagnets (FMs), the only viable way of switching their magnetization with light involves inducing thermal melting with strong optical pulses~\cite{Alebrand_APL_2012,Lambert_Science_2014,Gorchon_APL_2017}. The demand for the development of less-invasive optical orientation schemes is fueled by growing research interest in topological ferromagnets~\cite{Yu_Science_2010,Chang_Science_2013}, for which optical creation of magnetic domains would imply dynamic generation of local Chern domains: a feat that thus far could only be achieved with static methods~\cite{Yasuda_2017,Awschalom_2017,Rosen_2017}. Recently, a highly-tunable platform for accessing such ferromagnets has emerged with the advent of semiconducting moir\'e materials (SMMs), such as twisted MoTe$_2$ (t-MoTe$_2$) bilayers, that host flat, topologically-nontrivial valence bands~\cite{Wu_PRL_2019,Li_PRS_2021,Magic_Nat_Comm_2021}. This paved the way for the realization of electrically-tunable topological ferromagnetic metals as well as zero-magnetic-field fractional Chern insulators (FCIs)~\cite{Neupert_PRL_2011,Bernevig_PRX_2011,Long_Nature_Nanotech_2024} -- spin-polarized lattice analogs of fractional quantum Hall (FQH) states that feature topological-order and quasiparticle excitations with fractional statistics~\cite{Klitzing_PRL_1980,Schrieffer_PRL_1984,Klitzing_RMP_1986}, making them promising for novel quantum technologies~\cite{Fu_Kane_PRL_2008,Mellnik_Nature_2014}. 

\begin{figure*}[t]
	\includegraphics[width=0.99\textwidth]{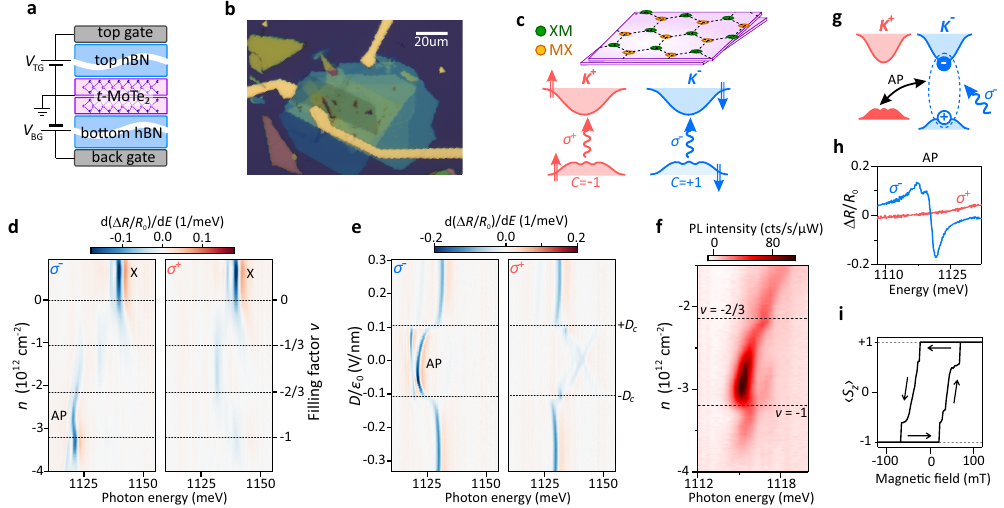}
	\caption{{\bf Attractive polaron as a resonant spin sensor of correlated topological phases.} ({\bf a}) Sketch and ({\bf b}) optical micrograph of device A. It consists of a t-MoTe$_2$ bilayer with a target twist angle of 3.5$^\circ$ that is embedded between two hexagonal boron-nitride (hBN) flakes and further sandwiched between two transparent graphene gates. ({\bf c}) Schematic illustration of an effective honeycomb moir\'e superlattice potential for layer-hybridized holes, which gives rise to flat topological valence bands with finite Chern numbers $C=\mp1$ in $K^\pm$ valleys that couple to $\sigma^\pm$-polarized light. ({\bf d},{\bf e}) Differentiated reflectance contrast spectra $\mathrm{d}(\Delta R/R_0)/\mathrm{d}E$ detected in two circular polarizations as a function of $\nu$ at fixed $D\approx0$ ({\bf d}) and a function of $D$ at fixed $\nu\approx-1$ ({\bf e}). The spectra were measured at a small magnetic field $B\approx50$~mT required to initiate the spin orientation of newly-injected holes upon changing the gate voltages. ({\bf f}) Doping-density evolution of non-resonantly excited PL spectra measured at $D\approx0$. The cusps in the trion emission energy are due to ICI and FCI formation. ({\bf g}) Cartoon illustrating spin-selective nature of exciton-polaron dressing, underlying locking between the hole spin orientation and circular polarization of the AP resonance. ({\bf h}) Circular-polarization-resolved $\Delta R/R_0$ spectra at $\nu=-1$ showing completely polarized AP resonances, which implies ferromagnetic order in the hole spin system. ({\bf i}) Hysteresis curve of measured at $\nu=-1$.} \label{fig:Fig1}
\end{figure*}

Here, we demonstrate all-optical control over the spin-valley orientation and many-body Chern number of strongly correlated topological phases in t-MoTe$_2$. Our method employs resonant, circularly-polarized excitation, which drives the many-body spin system between two degenerate metastable ground states with opposite Chern numbers. We evidence that topological ferromagnets remain incompressible while their Chern number is optically flipped, indicating that our orientation scheme does not involve light-induced heating. Our spatially-resolved experiments demonstrate that by choosing the strength and the spatial profile of the light beam, it is possible to write simple Chern domains and topological edge states that retain their shape after the excitation is turned off. This could pave the way for the development of optically-programmable topological circuits of arbitrary geometries, thus marking a breakthrough in the emerging field of topotronics~\cite{Hasan_Kane_RMP_2010,Gilbert_ComPhys_2021}.

Our experiments are carried out on two different charge-controlled t-MoTe$_2$ devices (Figs~\ref{fig:Fig1}{\bf a,b}) with intended twist angles of around 3.5$^\circ$ (device A) and 4.1$^\circ$ (device B). Unless otherwise stated, the results discussed in the main text are obtained for device A. As proven by previous studies~\cite{Zeng_Nature_2023,Park_Xu_2023,Xu_PRX_2023,Cai_Xu_2023}, t-MoTe$_2$ features strong inter-layer hybridization of valence bands~\cite{Anderson_Science_2023,Cai_Xu_2023,Zeng_Nature_2023,Xu_PRX_2023,Park_Xu_2023,Wang_PRL_2024} resulting in formation of a honeycomb moir\'e lattice (Fig.~\ref{fig:Fig1}{\bf c}) that supports flat valence bands with valley-contrasting Chern numbers $C=\mp1$. The hole density $n$ in these bands, and the corresponding moir\'e filling factor $\nu$, are set by the chemical potential $V_\mu$ that can be controlled independently of the displacement field $D$ using two bias voltages applied to top and bottom gate electrodes. To suppress thermal fluctuations, all our measurements were conducted in a dilution refrigerator unit with free-space optical access and electronic temperature around $\sim0.1$--$0.2$~K.


\begin{figure*}[t]
	\includegraphics[width=0.99\textwidth]{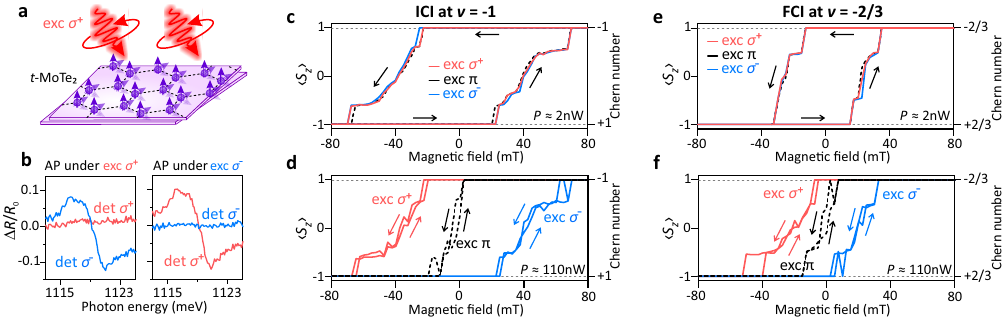}
	\caption{{\bf Controlling topological Chern number with light.} ({\bf a}) Cartoon illustrating the optical spin orientation with circularly polarized light. ({\bf b}) AP reflectance spectra measured at $\nu=-1$ and $D\approx0$ in two circular polarizations with weak ($\lesssim2$~nW) probe light just after exciting the spin system with $\sigma^+$ (left) and $\sigma^-$ (right) polarized beam at $P_\mathrm{pump}=110$~nW.  ({\bf c-f}) Magnetic hysteresis loops measured for the ICI at $\nu=-1$ ({\bf c,d}) and FCI at $\nu=-2/3$ ({\bf e,f}) using three different polarizations of excitation ($\sigma^\pm$ and linear $\pi$) at two different pump powers: 2~nW ({\bf c,e}) and 100~nW ({\bf d,f}). The collapse of bistable hysteretic behavior in the latter case demonstrates the feasibility of all-optical control over the spin polarization and Chern number of both strongly correlated topological phases.} \label{fig:Fig2}
\end{figure*}

Fig.~\ref{fig:Fig1}{\bf d} shows an example circular-polarization-resolved $n$-dependence of white light reflectance contrast $\mathrm{d}(\Delta R/R_0)/\mathrm{d}E$ spectrum (differentiated with respect to the photon energy $E$) obtained at vanishing $D\approx0$. In these measurements, the excitation power is set below 10~nW to suppress light-induced spin depolarization~\cite{Smolenski_PRL_2022,Ciorciaro_Nature_2023,Hongkun_Park_Nat_Phys_2025} and ensure that the incident light constitutes a non-destructive probe of the electronic spin system. Upon injection of itinerant holes, the excitons (X) in the two MoTe$_2$ layers are dynamically screened, giving rise to attractive polaron (AP) resonances that become visible at the onset of hole doping. Despite strong layer-hybridization of the hole wavefunctions, the excitonic resonances remain non-hybridized, which results in the presence of two pairs of distinct X and AP transitions, whose energies are likely lifted by a residual strain in the sample.

In stark contrast to the excitons, the AP transitions are visible only in one circular polarization for $|\nu| \gtrsim 0.4$. This is a consequence of spin-selective exciton-polaron dressing~\cite{Back_PRL_2017}, which stems from the fact that $\sigma^\pm$-polarized excitons in $K^+$ ($K^-$) valley interact attractively only with spin-down (spin-up) holes residing in the opposite $K^-$ ($K^+$) valley (Fig.~\ref{fig:Fig1}{\bf g}). As a result, the spectral weight $f_\mathrm{AP}^\pm$ of the AP resonances in $\sigma^\pm$ polarization is directly proportional to the hole density $n_\mp$ in the $K^\mp$ valley, rendering APs perfect optical spin sensors~\cite{Smolenski_PRL_2022}. Unlike previously-employed trion resonances accessible in photoluminescence (PL) spectra of t-MoTe$_2$~\cite{Anderson_2024}, the resonant excitation of AP resonances we use here does not involve an intermediate energy relaxation process; this ensures that the degree of AP circular-polarization $\rho_\mathrm{AP}=(f_\mathrm{AP}^--f_\mathrm{AP}^+)/(f_\mathrm{AP}^-+f_\mathrm{AP}^+)$ corresponds precisely to the degree of spin-valley polarization $\langle S_z\rangle=(n_+-n_-)/(n_++n_-)$ of the hole system. 

In view of the above considerations, the complete AP polarization we observe (Fig.~\ref{fig:Fig1}{\bf h}) indicates the full hole spin polarization for $|\nu| \gtrsim 0.4$. This is expected for ferromagnetic phases, whose formation manifests in a prominent magnetic hysteresis loop (Fig.~\ref{fig:Fig1}{\bf i}). Such hysteresis appears in a broad range of filling factors $-1.3\lesssim\nu\lesssim-0.4$, covering both ferromagnetic metals as well as integer and fractional Chern insulators (ICIs and FCIs) at $\nu=-1$ and $\nu=-2/3$, respectively~\cite{Anderson_Science_2023,Cai_Xu_2023,Zeng_Nature_2023,Xu_PRX_2023,Park_Xu_2023}. The formation of the latter gapped phases gives rise to characteristic cusps in the energies of optical resonances, as seen for the AP around $\nu=-1$ in the resonant reflectance spectra (Fig.~\ref{fig:Fig1}{\bf d}) and, more clearly, for the trion in PL spectra around both $\nu=-1$ and $\nu=-2/3$ (Fig.~\ref{fig:Fig1}{\bf f}). Consistent with previous reports~\cite{Park_Xu_2023,Cai_Xu_2023,Zeng_Nature_2023,Xu_APS_2023}, we find that hole densities corresponding to these cusps increase linearly with external magnetic field according to the Streda formula, confirming the topological nature of both insulating phases. Their stability can be controlled with the displacement field $D$ that suppresses interlayer hybridization, and eventually drives a topological phase transition to trivial bands beyond critical $|D|>D_c$, manifesting in the loss of circular polarization of the optical resonances (Fig.~\ref{fig:Fig1}{\bf e})~\cite{Cai_Xu_2023,Anderson_Science_2023,Wang_PRL_2024}.


\begin{figure*}[t]
	\includegraphics[width=0.99\textwidth]{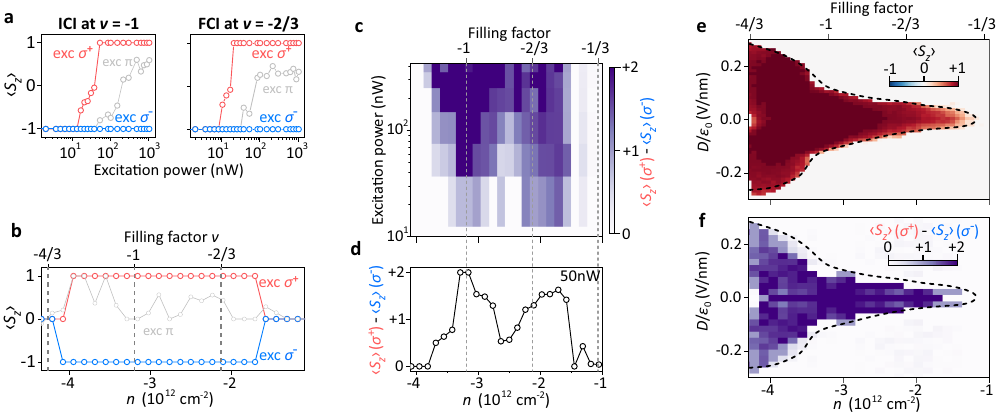}
    \caption{{\bf Efficiency and robustness of the optical spin orientation.} ({\bf a,b}) Excitation power ({\bf a}) and electron density ({\bf b}) dependence of the hole spin polarization induced by circularly- and linearly-polarized light. For each data point, the spins are first oriented downwards by applying $B=-50$~mT, which is then turned off. The data are obtained at $D\approx0$. ({\bf c})~Color-scale plot showing excitation power and hole density dependence of the optical spin orientation efficiency defined as a difference between $\langle S_z\rangle$ under $\sigma^+$ and $\sigma^-$-polarized excitation.
    ({\bf d}) Linecut through ({\bf c}) at $P_\mathrm{pump}=50$~nW.  ({\bf e}) Color-scale map showing the remanent hole spin polarization degree measured with weak linearly-polarized probe light upon ramping the field from $50$~mT to 0. The black dotted line marks the phase space where the holes are ferromagnetically ordered. ({\bf f}) A corresponding color-scale map showing the optical spin orientation efficiency $\langle S_z\rangle(\sigma^+)-\langle S_z\rangle(\sigma^-)$ at $P_\mathrm{pump}=300$~nW and $B=0$ (note that the data for $D<0$ is obtained by mirroring the measured data for $D>0$). The black dotted line is the same as in ({\bf e}).} \label{fig:Fig3}
\end{figure*}

Having established direct resonant readout of the hole spin state in t-MoTe$_2$, we proceed with implementing optical spin-valley orientation. Our approach is based on illuminating the sample with circularly-polarized photons, whose angular momentum is transferred to the resident holes (Fig.~\ref{fig:Fig2}{\bf a}). To this end, we first narrow down the excitation bandwidth to a $\sim5$~meV-wide spectral window centered around the AP resonance energy. We optically pump the hole spin system for a few seconds (at a fixed $\nu$, $D$, and $B$) with a given power $P_\mathrm{pump}$ and a selected polarization (circular or linear). To probe the efficiency of optical orientation, we then set the excitation polarization to be linear and reduce the power down to $\lesssim2$~nW, which does not influence the spin-valley degree of freedom of the hole system. This simple pump-probe scheme allows us to directly sense the resulting hole spin orientation by analyzing the AP spectral weights in $\sigma^\pm$-polarized components of the reflected light.

Figs.~\ref{fig:Fig2}{\bf c,d} show magnetic hysteresis measured in such a way for the ICI at $\nu=-1$ at two different pump powers for linearly ($\pi$) and circularly-polarized ($\sigma^+$ and $\sigma^-$) pump fields. As stated above, optical pumping does not influence the hysteresis loop as long as $P_\mathrm{pump}$ does not exceed a few nW (Fig.~\ref{fig:Fig2}{\bf c}). This, however, changes drastically when $P_\mathrm{pump}$ is increased beyond 100~nW (Fig.~\ref{fig:Fig2}{\bf d}): in this case, bistable hysteretic behavior almost completely disappears, and the hole spin orientation becomes uniquely determined by the optical pump polarization and the magnetic field. In particular, when $B$ remains lower than the original coercive field $B_c\approx50$~mT, all hole spins within the excitation spot can be optically oriented from down to up by switching the helicity of circular polarization of the pump from $\sigma^-$ to $\sigma^+$ (and vice versa). This indicates that the Chern number of the hole system changes from $C=+1\leftrightarrow C=-1$, as all the holes get transferred to the optically-driven valley, resulting in the formation of a dark steady-state of the driven-dissipative many-body system. This is further confirmed by the disappearance of the AP resonance in the circular polarization used for optical pumping (see Fig.~\ref{fig:Fig2}{\bf b}). This process becomes inefficient beyond $|B|>B_c$, when the spin orientation is set by the magnetic field direction. As expected, upon switching the pump polarization to linear, the hole spin orientation becomes effectively randomized, bringing the critical magnetic field required to induce a spin-flip close to 0 (a slight offset from $B=0$ observed under linear excitation in  Figs.~\ref{fig:Fig2}{\bf d,f} is attributed to the remanent field of our superconducting magnet).

Crucially, the optical Chern number switching scheme we developed is not limited to the ICI, but works equally well for the FCI at $\nu=-2/3$. This is demonstrated by the power- and polarization-dependent magnetic hysteresis measurements in Figs.~\ref{fig:Fig2}{\bf e,f}. This observation provides the first evidence for all-optical control of a topologically-ordered phase of matter.

\begin{SCfigure*}[][t!]
    \includegraphics{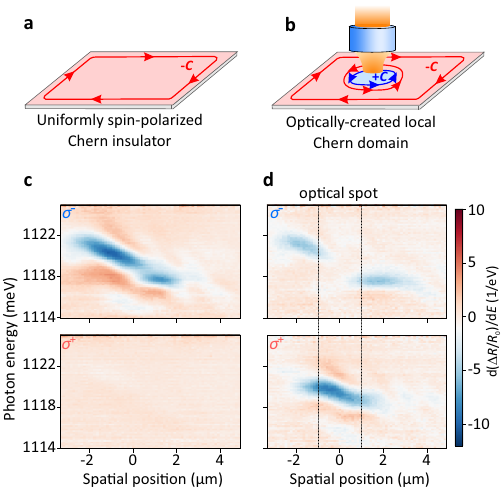}
	\caption{{\bf Optical writing of topological edge modes.} ({\bf a,b}) Schematic illustrating all-optical writing of topological Chern junctions: by illuminating selected area with circularly-polarized light, we can define a local Chern domain, which is surrounded by a chiral topological edge mode. ({\bf c,d}) 1D spatial maps showing differentiated AP reflectance contrast spectra obtained at $B=0$, $\nu=-1$, and $D\approx0$, in two circular polarizations of detection under weak, linearly-polarized probe light. This data were obtained in device B owing to its larger spatial homogeneity. Before acquiring ({\bf c}), the spins across the entire sample are oriented upwards with the external magnetic field. Then, after measuring ({\bf c}), the spin system is optically oriented with 70nW of $\sigma^-$-polarized light focused to a narrow spot in the central region of the sample (the extent of which is marked by vertical dashed lines). Finally, ({\bf d}) is acquired. The coexistence of regions with opposite AP polarization evidences the formation of topological Chern junction.}
	\label{fig:Fig4}
\end{SCfigure*}

To further examine the efficiency of the spin pumping method, we perform excitation-power dependent experiments for both ICI and FCI phases (Fig.~\ref{fig:Fig3}{\bf a}). In these measurements, the spins are first oriented downwards using a magnetic field $B<-B_c$, which is then ramped to 0, where the optical orientation is explored. Clearly, the orientation exhibits non-linear dependence on the pump power: to optically switch the spins from $\langle S_z\rangle=-1\rightarrow+1$, the power has to exceed a certain threshold that is slightly larger for the ICI, but is generally around a few tens of nW.

Our method turns out to be extremely robust, allowing us to optically orient the spins in both insulating and metallic phases both above and below $\nu=-1$ (Fig.~\ref{fig:Fig3}{\bf b}). Interestingly, the critical power level necessary to orient the metallic phase around at $\nu\approx-0.8$ is considerably higher than that for the Chern insulators at $\nu=-1$ and $\nu=-2/3$ (Fig.~\ref{fig:Fig3}{\bf c,d}). 

At large enough powers $P_\mathrm{pump}\approx 300$~nW, the spin pumping remains efficient in the almost entire $(\nu,D)$ phase space where the hole system is ferromagnetic. This is directly revealed by Figs.~\ref{fig:Fig3}{\bf e,f} showing a comparison between the remanent hole spin polarization $\langle S_z\rangle$ obtained at low optical power upon ramping $B$ from $B_c$ to 0, and the difference between $\langle S_z\rangle$ values that can be reached at $B=0$ upon switching the pump polarization from $\sigma^-$ to~$\sigma^+$.

In view of the above findings, we speculate that the mechanism behind the investigated optical orientation process is based on optical injection of itinerant holes into the targeted, originally unoccupied valley. When all holes are initially in the $K^+$ valley, optical excitation of an AP in $\sigma^-$ polarization results in the generation of a bound electron-hole pair in the $K^-$ valley. While its radiative recombination would leave the hole spin state unchanged, a spin relaxation event related to AP inter-valley scattering~\cite{Glazov_PRB_2019} or donor-assisted recombination~\cite{Dery_PRL_2014} might effectively result in the transfer of a hole to the $K^-$ valley. Even if the probability of such an event is small, continuous optical drive will lead to the formation of metastable domains consisting of holes in the $K^-$ valley, provided that their spin relaxation back to the lower-energy $K^+$ valley is sufficiently slow. Given that this relaxation will further slow down as optically created domains grow larger, this process will eventually result in flipping the spin-valley index of the entire hole system. It is remarkable that, despite involving multiple incoherent photon/phonon emission events, this mechanism does not lead to light-induced melting of the Chern insulating phases, which remain robust under the requisite excitation power levels.


A key advantage of our approach lies in its inherently local character, enabling switching of the many-body Chern number within a micron-sized region defined by the optical spot. This opens up an exciting opportunity to optically create Chern junctions and write topological edge modes in an arbitrarily selected area of the device, away from its physical edges, as schematically illustrated in Figs.~\ref{fig:Fig4}{\bf a,b}.

To demonstrate this, we perform spatially-resolved experiments, in which we monitor the effect of local optical pumping of the ICI at $\nu=-1$ on the hole spin polarization in remote spots that are distributed along a selected horizontal direction. We first ensure that the holes in the entire device are in $C=-1$ state by applying an external magnetic field. Then, the field is lowered down to 0; at this point, the AP resonance is visible exclusively in $\sigma^-$ polarization in all investigated locations (Fig.~\ref{fig:Fig4}{\bf c}) despite clear spatial inhomogeneities inherent to van der Waals heterostructures, which manifest in sizable spatial variations of the optical resonance energy. Next, we optically orient the hole spins in a given, centrally-located spot by resonantly driving it for a few seconds with $\sigma^-$-polarized light resonant with the AP transition at $P_\mathrm{pump}\approx70$~nW. Finally, a weak linearly-polarized light is exploited to readout the resulting spatial distribution of the hole spin orientation by mapping out the polarization degree of the AP resonance (Fig.~\ref{fig:Fig4}{\bf d}). 

As expected, the spins within the optically-pumped spot are flipped, as evidenced by the reversed helicity of the AP resonance. The original AP polarization remains, however, not affected in all adjacent regions. This demonstrates all-optical generation of a Chern junction between a local magnetic domain with $C=+1$ and oppositely-magnetized regions with  $C=-1$. This observation implies the generation of chiral topological edge modes that support unidirectional current flow at the interfaces between these regions. Crucially, we find that a similar approach enables writing of the edge modes for the FCI state at $\nu=-2/3$. Together with the ability to optically reposition the Chern junction by either shifting the optical pumping spot or simply increasing the excitation power, our approach paves the way for writing of optically-programmable topological circuits. Successful realization of this concept will, however, require improvements in the spatial homogeneity of t-MoTe$_2$ based devices, as local disorder and twist angle variations in the present samples often limit the size of the optically-generated Chern domains.


Our method of controlling the spin-valley degree of freedom of Chern insulators and ferromagnetic metals opens up novel possibilities for all-optical manipulation of topological phases of matter. In particular, understanding the dynamics of the formation and growth of optically-created spin-valley-flipped domains constitutes an exciting research direction. Another interesting extension of our work would be to employ ac Stark shift induced by short, circularly-polarized optical pulses red-detuned from the AP transition~\cite{Uto_PRL_2024,PRX_Bertrand}, to generate effective pseudomagnetic fields~\cite{Awschalom_1999}. For sufficiently high pulse energies, this approach might enable ultra-fast flipping of the topological Chern numbers without the need for creating real optical excitations in the system. 

A unique aspect of the optical orientation we demonstrated is that it allows for controlled generation and manipulation of topological edge modes. In the future, by combining optical writing with electric transport, it might be possible to create edge modes of any conceivable geometry, for example enabling explorations of anyon interferometry and edge state tunneling.

\subsection*{Acknowledgments}
This work was supported by the Swiss National Science Foundation (SNSF) under Grant Number 200021-204076. E.A., W.L., and X.X. were supported by the US DOE BES (DE-SC0012509) and Vannevar Bush Faculty Fellowship (Award number N000142512047). K.W. and T.T. acknowledge support from the JSPS KAKENHI (Grant Numbers 21H05233 and 23H02052), the CREST (JPMJCR24A5), JST and World Premier International Research Center Initiative (WPI), MEXT, Japan. We thank Clemens Kuhlenkamp, Michael Knap, Haydn Adlong, Arthur Christianen and Liu Zheng for insightful discussions.

\subsection*{Author contributions}
T.S. designed the experiments and carried out initial observations. T.S. and A.I. supervised the project. O.H., K.K., and T.S. performed the measurements and analyzed the experimental data with assistance from M.K., E.A., W.L., and X.X. W.L. and E.A. fabricated the samples. K.W. and T.T. provided bulk hBN crystals. T.S., O.H., and A.I. wrote the manuscript, with input from all authors.

\bibliographystyle{apsrev_my_3}
\bibliography{references}

\end{document}